\documentstyle[12pt,twoside,fleqn,amssymb,epsfig,espcrc1,graphicx]{article}
\newdimen\htxtw
\htxtw=\textwidth
\divide\htxtw by 2

\title{Recent results from Pb-Au collisions at
  158 GeV/c per nucleon \\ obtained with the CERES spectrometer}

\author{
B.~Lenkeit for the CERES Collaboration:\\ \medskip
G.~Agakichiev$^b$,
H.~Appelsh\"auser$^e$,
R.~Baur$^e$,
P.~Braun-Munzinger$^a$,
F.~Ceretto$^{d,*}$,
A.~Cherlin$^f$,
A.~Drees$^{e,*}$,
S.~Esumi$^e$,
U.~Faschingbauer$^{e, d}$,
Z.~Fraenkel$^f$,
P.~Gl\"assel$^e$,
G.~Hering$^a$,
P.~Holl$^g$,
Ch.~Jung$^e$,
B.~Lenkeit$^e$,
A.~Marin$^a$,
M.~Messer$^e$,
D.~Miskowiec$^a$,
Y.~Panebrattsev$^b$,
A.~Pfeiffer$^e$,
J.~Rak$^d$,
I.~Ravinovich$^f$,
S.~Razin$^b$,
P.~Rehak$^g$,
M.~Richter$^e$,
N.~Saveljic$^b$,
W.~Schmitz$^e$,
J.~Schukraft$^c$,
W.~Seipp$^e$,
S.~Shimanskiy$^b$,
J.~Slivova$^d$,
E.~Socol$^f$,
H.J.~Specht$^e$,
J.~Stachel$^e$,
I.~Tserruya$^f$,
Th.~Ullrich$^e$,
C.~Voigt$^e$,
S.~Voloshin$^h$,
C.~Weber$^e$,
J.P.~Wessels$^e$,
T.~Wienold$^e$,
J.P.~Wurm$^d$,
V.~Yurevich$^b$ \\ \medskip
$^a$ GSI, Darmstadt, Germany \\
$^b$ JINR, Dubna, Russia \\
$^c$ CERN, Geneva, Switzerland \\
$^d$ Max-Planck-Institut f\"ur Kernphysik, Heidelberg, Germany \\
$^e$ Universit\"at Heidelberg, Germany \\
$^f$ Weizmann Institute of Science, Rehovot, Israel \\
$^g$ Brookhaven National Laboratory, Upton, USA \\
$^h$ Guest at Universit\"at Heidelberg, Germany \\
$^*$ now at State University New York at Stony Brook, USA \\
}

\begin{document}
\maketitle

\begin{abstract}
During  the 1996 lead run time, CERES has accumulated
42 million events, corresponding to a factor of 5 more statistics than in 1995 and
2.5 million events of a special photon-run. We report on the results of the
low-mass e$^+$e$^-$-pair analysis. Since the most
critical item is the poor signal-to-background ratio we also discuss the 
understanding of this background, in absolute terms,
with the help of a detailed Monte Carlo simulation. 
We show preliminary results of the photon analysis and summarize
the results of the hadron analysis preliminarily reported on already at QM'97 
\cite {fede_qm97}.
\end{abstract}
  
\section{Introduction}
Since electromagnetic probes have a mean free path which is much larger than
the size of the reaction volume of a heavy ion collision, they carry
direct information also from the earliest stages of the collision. The CERES
spectrometer is optimized to measure low-mass e$^+$e$^-$-pairs, including
the low pair-p$_\perp$-region which was not accessible in previous dimuon experiments
at CERN. Also photons can be detected by the conversion method \linebreak ($\gamma$ 
$\rightarrow$ e$^+$e$^-$). The main advantage 
of dilepton pairs over photons is that they carry
invariant mass which helps to distinguish between different production
processes, and in particular to observe vector mesons by their direct decay. \hfill\break
In order to extract signals from the early stages of the collision, detailed 
knowledge of the electromagnetic decays of hadrons
after freezeout is necessary. Such a reference was provided by
measuring the inclusive pair production in p-Be collisions \cite {pBe}. Together
with an electromagnetic calorimeter (TAPS), we have exclusively reconstructed the 
$\pi^0$- and $\eta$-Dalitz decays. It was found that the inclusive pair
yield per charged particle 
can be fully understood by hadron decays in p-Be collisions as well
as in p-Au collisions. It is with respect to this reference that CERES 
has observed
an enhanced pair production, first in S-Au collisions \cite {SAu} and later also
in Pb-Au collisions \cite {PbAu}. 
The inclusive photon analysis of the S-Au data did not show a significant 
excess compared to the expectations from hadron decays \cite {SAu-gamma}.\hfill\break
\section{Experimental Setup}
A main experimental problem a dilepton experiment in nucleus-nucleus 
collisions has to deal with, is the
small amount of leptons compared to the huge number of charged hadrons
(N$_{e^+e^-}$/N$_{ch}$ $\sim$ 10$^{-5}$). This problem is attacked by using
two Ring Imaging CHerenkov detectors with a  
$\gamma_{th}$ = 32, such that 95\%
of all charged hadrons do not produce Cherenkov 
\begin{figure}[ht]
    \vskip -1cm
    \hspace*{-5cm} 
    \begin{center} 
    \mbox{\epsfig{file=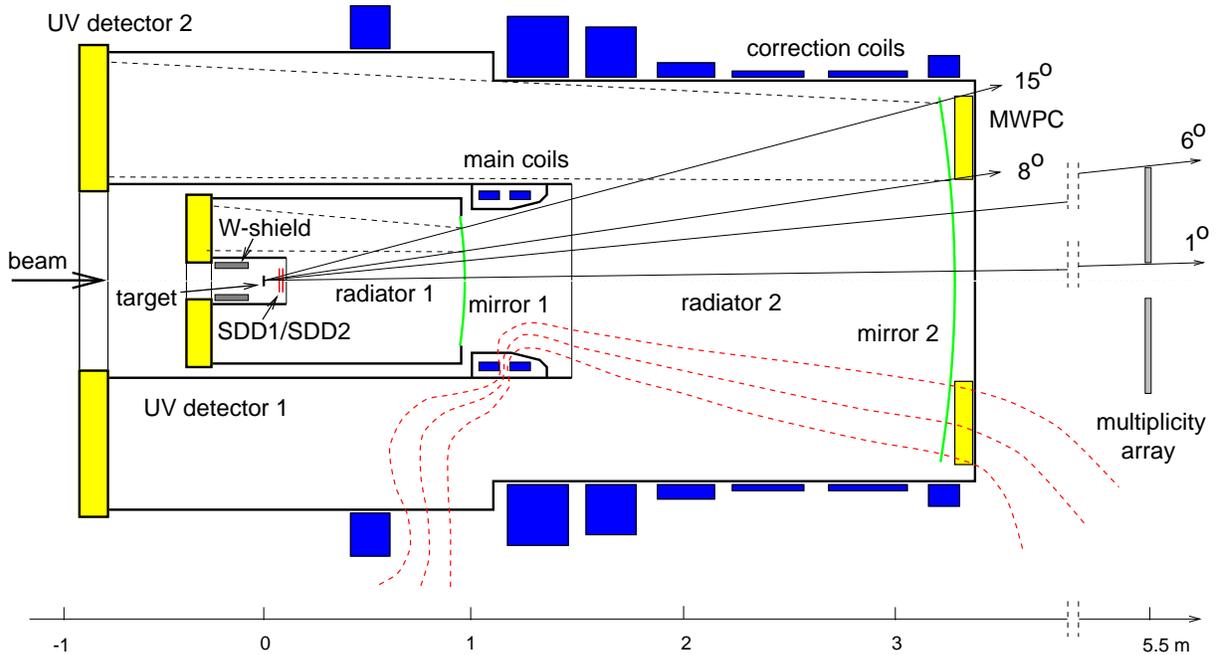,width=1.\textwidth}}
    \end{center}
    \vskip -1.cm
    \caption
    {Experimental setup of the CERES experiment during the Pb data taking. The 
    dashed curves show the axially-symmetric magnetic field lines.}
    \label{setup}
\end{figure} 

\begin{figure}[ht]
     \vskip -0.5cm
\centering
\resizebox{\textwidth}{!}{\rotatebox{270}{%
\includegraphics*[bb=44 2 483 797]{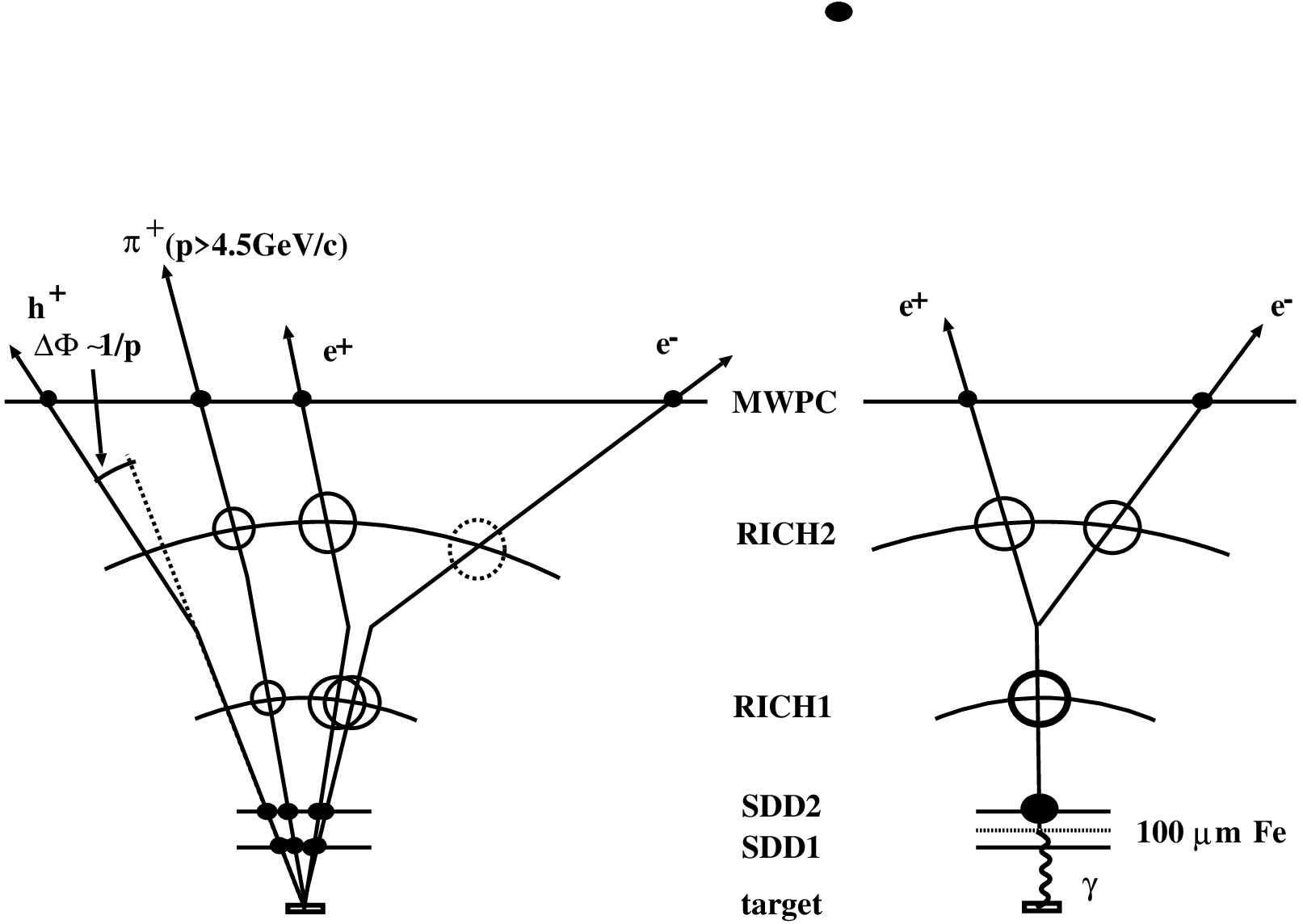}}}
      \vskip -0.9cm  
\caption{\label{trackpattern} Signature of different physics sources in the 
        CERES spectrometer. \hfill\break Left Side:
        Hadrons with a momentum below the Cherenkov 
        threshold (labelled h$^+$) are only visible in the external tracking detectors
        (SDD's and MWPC). High
        momentum pions (labelled $\pi^+$) and electrons 
        (labelled e$^+$ and e$^-$) are also visible in the 2 RICH's
        and can be distinguished by
        the ring radius. The two electron tracks show the typical signature
        of a $\pi^0$-Dalitz decay where the matching ring in RICH2 is not found; two incomplete pairs like
        that are the main source of combinatorial background. The genuine high mass signal
        pairs we are interested in have two well sepaprated tracks already in the first detectors
        (not contained in the figure).\hfill\break 
        Right side: The 
        signature of a photon in our special photon run is shown. Since here the background is
        negligible, it is not indicated.} 
\vskip -0.5cm
\end{figure}
\vskip -0.5cm
\hskip -0.4cm 
light. As shown in Fig.~\ref {setup}, these RICH detectors are assisted by three external tracking
detectors, two Silicon Drift Detectors (SDD1,2) 10 cm behind the target and a Multi-Wire
Proportional Chamber (MWPC) at the end of the spectrometer. The signatures
of charged-particle tracks are shown in Fig.~\ref {trackpattern}. While the 95 \% of all 
charged hadrons
below the Cherenkov threshold are only visible in the external tracking detectors,
the pions with a momentum larger than 4.5 GeV/c are also visible in the 
RICH detectors where they produce a ring the radius of which depends on the 
relativistic $\gamma$ factor of 
the pion and is significantly smaller than the asymptotic radius. Therefore,
pion rings are well distinguished from electron/positron rings, which always have
asymptotic radii.\hfill\break
The large number of photons per e$^+$e$^-$-pair, also of order 10$^{5}$ demanded
to strictly minimize the radiation length within the detector acceptance (X/X$_0$ $<$ 1 \%),
so that only ~1\% of the photons convert into a e$^+$e$^-$-pair. 
This is
still, together with the decay $\pi^0$ $\rightarrow$ e$^+$e$^-\gamma$, the
do\-mi\-nant source for dielectron pairs, both producing pairs with a mass smaller
than typically 200 MeV/c$^2$ (N$_{e^+e^-}$(m$>$200MeV/c$^2$ )/N$_{e^+e^-}$(m$<$200MeV/c$^2$ ) 
$\sim$ 10$^{-4}$).
Since it is impossible to observe on top of such sources an additional
source from an earlier stage of the collision, we are mainly interested in pairs 
with mass larger 200 MeV/c$^2$.\hfill\break
Because of our limited track reconstruction efficiency and our limited acceptance,
it occurs very often that from a physical pair only one leg is reconstructed
(as indicated in Fig.~\ref {trackpattern}).
Two of such partially reconstructed pairs in the same event produce a combinatorial 
background pair. To avoid this background a more efficient recognition especially
of pairs with mass smaller 200 MeV/c$^2$ is required.  
Therefore the third main idea
of the setup is to have three detectors 
(2 SDD's and RICH1) which are able to observe the original
small opening angle of these pairs which is later used for
background rejection. Therefore the shape of the magnetic field is such that
up to RICH1 the spectrometer is field-free, between RICH1 and RICH2 the 
charged particles get a deflection in azimuthal direction and afterwards the
field-lines point to the target (see Fig.~\ref {setup}), such that the particles have again straight 
tracks. The momentum
resolution due to detector resolution and multiple scattering is
$\Delta$p/p = ((2.3\% $\cdot$ p)$^2$ + (3.5\%)$^2$ )$^{1/2}$, with p in GeV/c.\hfill\break
For a 5 days running period we took data with a slightly changed setup: between 
the 2 SDD's we put a 100 $\mu$m steel foil as photon converter. Photons
have a very characteristic signature in this setup (see Fig.~\ref {trackpattern}):
they have no hit in the
first SDD,
a double minimum ionization pulse hight signal in the second SDD, a double
ring in the first RICH and two
well separated
rings in the second RICH accompanied by corresponding hits in the MWPC.
      
\section{Contributions from hadron decays}
As already mentioned the contributions of hadron decays to the dilepton and photon
yield can be well understood in proton-induced collisions.
They depend on
the yield of the hadrons, the decay branching ratio and, since we
have a limited kinematic and geometrical acceptance, also on the p$_\perp$ and 
rapidity distributions of the hadrons. Except for the branching ratios all other
ingredients are changing when going from proton-induced to lead-induced
collisions, as observed by several CERN-SPS experiments:
\begin {itemize}
\item The production ratios of the heavier hadrons are enhanced 
      as for example the $\eta$/$\pi^0$ 
      \cite{eta/pi} and the 2$\phi$/($\pi^+$+$\pi^-$) ratio
      \cite{phi/pi,phi/pi_2}. These particle ratios are best described in a thermal model, assuming
      chemical equilibrium \cite {chem-equ}, and presently unknown ratios
      have been extrapolated from it.
\item The inverse m$_\perp$-slope is systematically
      increasing with the mass of the hadron \cite {stachel_paris}, which is
      a hint for radial flow. The functional behaviour is: \\
      inverse m$_\perp$-slope (mass$_{hadron}$) = 0.175 MeV + 0.115 MeV/GeV $\cdot$ mass$_{hadron}$
\item While the widths of the hadron rapidity distributions 
      decrease with 
      particle mass in proton
      induced collisions, this is not observed in lead induced 
      collisions \cite{hohne}.
\end {itemize}  
All these observed changes are implemented in the generator simulating the
contributions of hadron decays and all the data shown in this paper are 
compared to the outcome of this generator. Compared to the pure p-p cocktail
used previously
this generator predicts a 30\% larger yield for m$_{e^+e^-}$ $>$ 0.2 GeV/c$^2$
(see Fig.~\ref {result1}). For the photon yield the correct description
of the $\pi^0$ distributions in our acceptance is most 
important, since the \linebreak $\pi^0$ $\rightarrow$ $\gamma\gamma$ decay dominates the 
yield (see Fig.~\ref {photon}).
      
\section{Low-mass e$^+$e$^-$-pair analysis}
The main challenge in the low-mass pair analysis is to reject the partially
recognized pairs with small opening angle (as indicated in Fig.~\ref 
{trackpattern}), while keeping a high reconstruction 
efficiency for pairs with large opening angle \cite {bibel}.
To understand the sources of background quantitatively, we have done a detailed MC simulation
using a generator for the composition of the sources and the decay kinematics.
The trajectories of these particles are  
traced through a GEANT simulation of our spectrometer. The resulting dE/dx and 
Cherenkov photon signal is then digitized in the same way as the data. 
The results are overlaid on real events, such that they can be 
analyzed in the same way as the data. Doing this for conversions and 
$\pi^0$-Dalitz decays on one hand and for pairs with masses larger
200 MeV/c$^2$ on the other hand, one obtains an objective tool to optimize the cuts.\hfill\break
\begin{figure}[ht]
  \begin{flushleft}
   \vskip -1.7cm 
    \begin{minipage}{\linewidth}
    \mbox{\hskip 0.5cm\epsfig{file=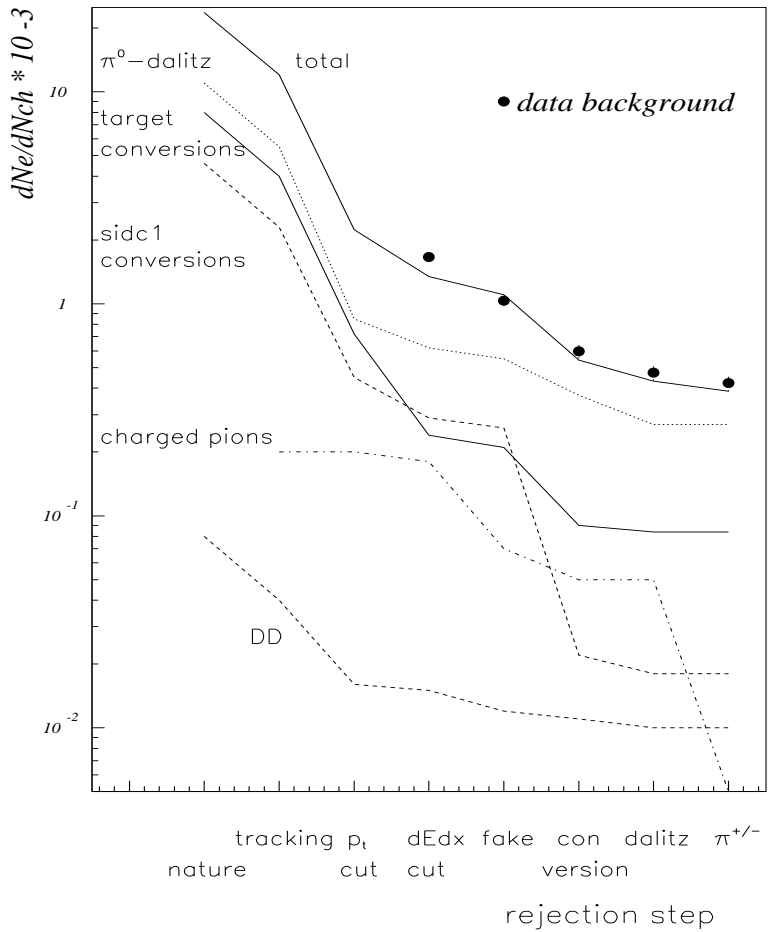,width=0.7\textwidth,
            height=0.83\textwidth}}
    \end{minipage}          
  \end{flushleft}
  \begin{flushright} 
  \vskip -13.8cm  
    \begin{minipage}{4.5cm}
    \caption{\label{compare_back}
    Relative number of tracks from various sources of combinatorial background
    shown along the sequence of rejection steps applied in the analysis.
    The points indicate the observed background in the data, the lines show 
    the estimates by the Monte Carlo Simulation.}
    \end{minipage}
  \end{flushright}  
  \vskip 5.4cm  
\end{figure}
\hskip -0.15cm
The rejection power of these cuts can be seen from Fig.~\ref {compare_back}
where the absolute contributions of all background sources are followed in chronological order 
over the different rejection steps. The most important rejection
steps are called "dE/dx" and "conversion" which use the property of the 
two SDD's to show a double amplitude for two very close tracks and 
"dalitz" which tries to identify close pairs by a close partner 
ring in RICH1. The MC simulation is compared with the observed
background in the data. The data follow very well the trend of the simulation
even in absolute terms.\hfill\break
At the end of all steps the signal-to-background ratio for pairs with mass
larger 200 MeV/c$^2$ is 1/13, with a mean reconstruction efficiency of 13\%.
The signal is extracted by
subtracting the number of like-sign pairs from the number of unlike-sign pairs.
By doing this one has to be sure that the combinatorial background
is equal in these two samples. This has been verified on the 
data by reversing all rejection cuts, such that one obtains a "clean" background
sample. This study indicates, that we can exclude an asymmetry of more than
1\% with a 90\% confidence level.
\subsection{Results}
\begin{figure}[ht]
  \vspace{-1.8cm}  
  \begin{flushleft}
    \begin{minipage}{0.5\linewidth}
      \epsfig{figure=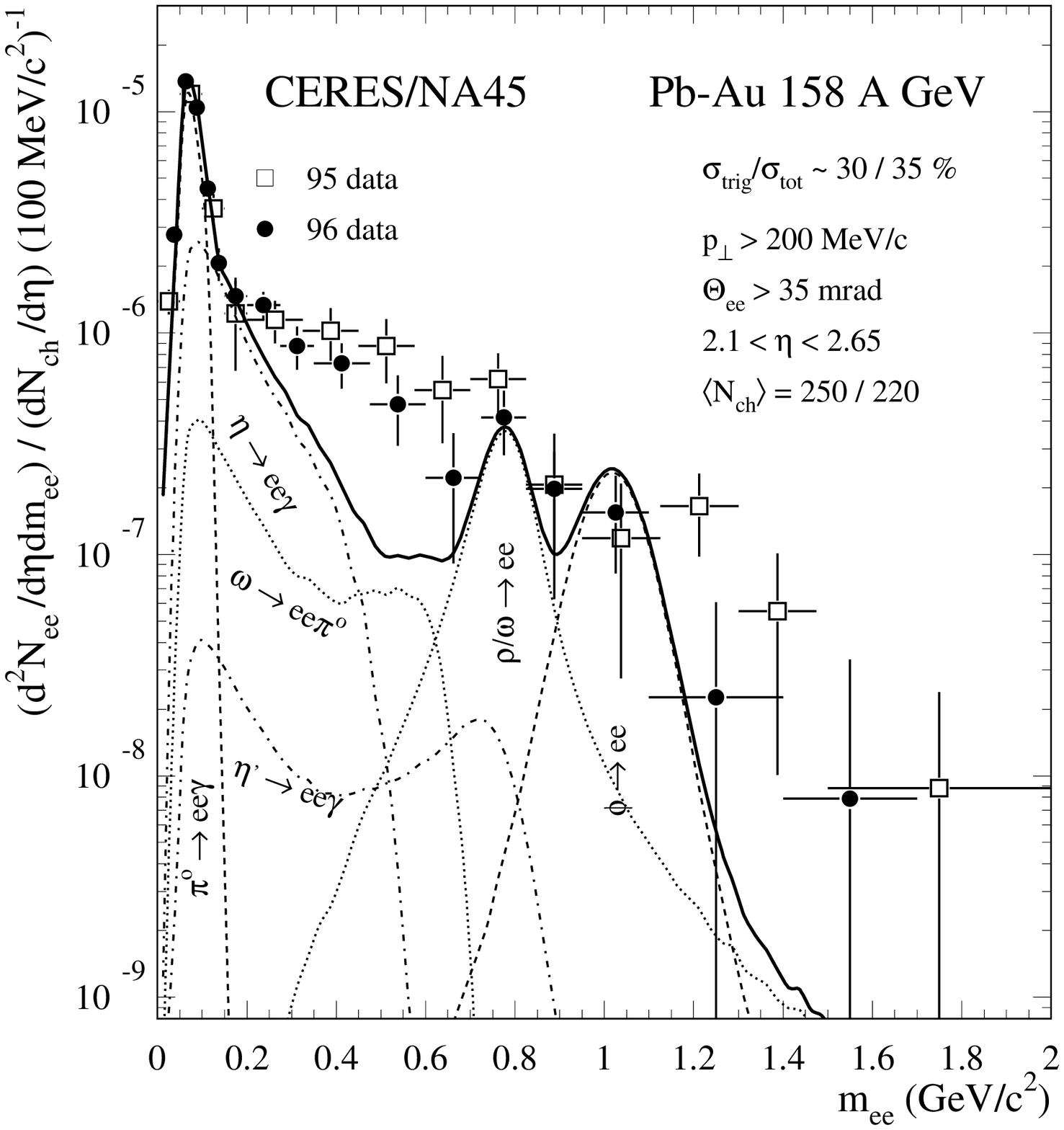,width=1.\linewidth}      
    \end{minipage}  
  \end{flushleft}
  \vspace{-8.2cm}
  \begin{flushright}
    \begin{minipage}{0.5\linewidth}
      \epsfig{figure=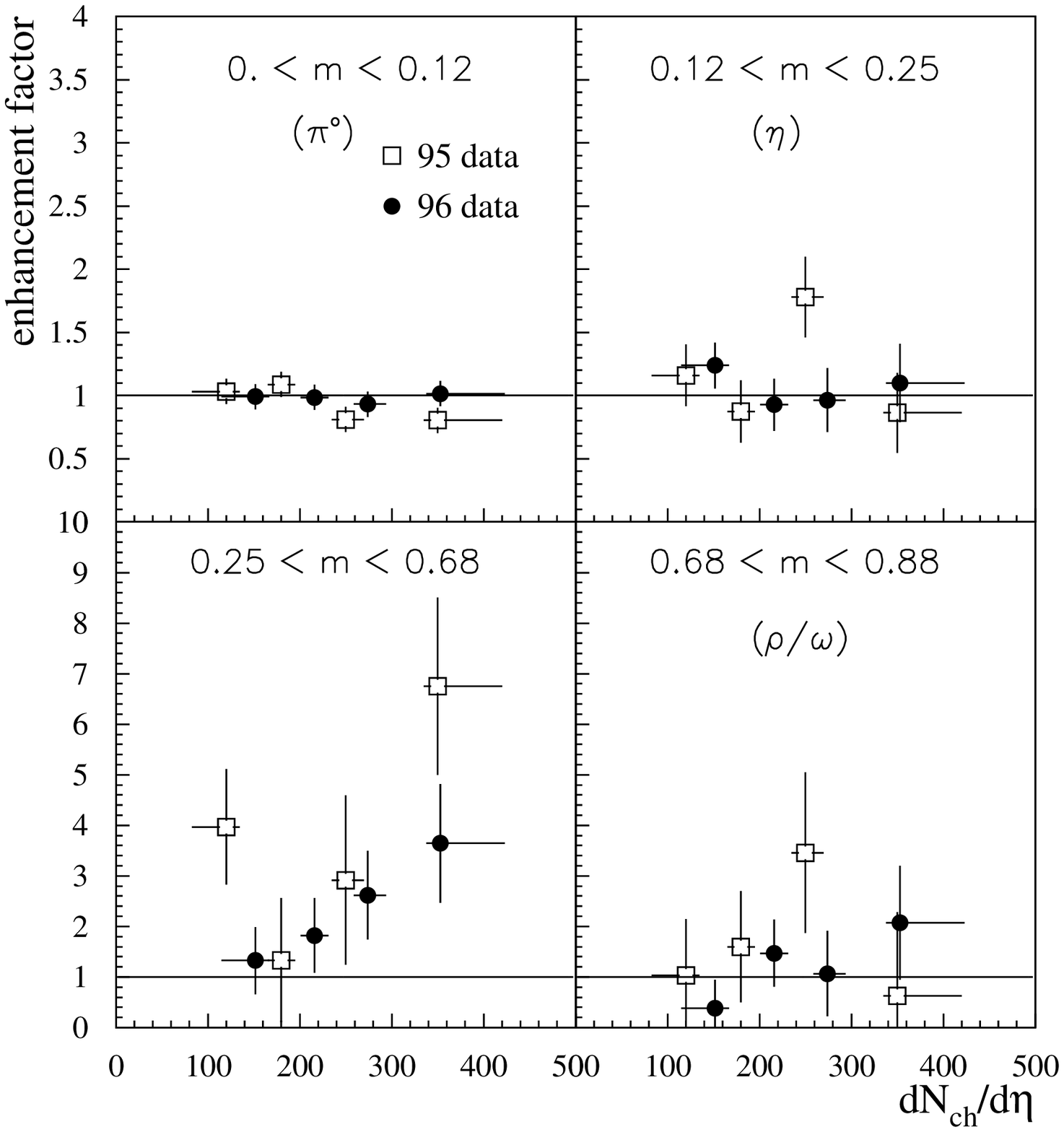,width=1.\linewidth}
    \end{minipage}
  \end{flushright}
  \begin{flushleft}
  \vspace{-3.5cm}    
\begin{minipage}{0.48\linewidth}
  \caption{   \label {result1}  Inclusive e$^+$e$^-$ mass spectrum obtained from the $'$95 
  and $'$96 data analyses. The data are compared to the sum of the expected contributions
  from hadron decays (solid line).}
\end{minipage}
  \end{flushleft}
  \vspace{-4.4cm}  
 \begin{flushright}
\begin{minipage}{0.48\linewidth}
  \caption{   \label {result2}Enhancement factor defined by
  N$^{e^+e^-}_{measured}$/N$^{e^+e^-}_{hadronic\ sources}$,
  within the mass range indicated
  as a function of the charged particle rapidity density
  of the collision.}  
\end{minipage}
  \end{flushright}
   \vspace{-0.1cm} 
\end{figure}

Fig. \ref {result1} shows the resulting invariant mass distribution of the observed
pair yield in comparison to the expectations from hadron decays. As 
previously reported by CERES we again observe an enhanced dilepton production
in the mass window starting around 200 MeV/c$^2$ up to the $\rho/\omega$ peak.
For comparison
the earlier results \cite {PbAu} are also shown. The centrality selection of the
  two data sets is slightly different, the mean number of charged particles
  per unit of rapidity is $<$N$_{ch}>$ = 250 and 220 for $'$96 and $'$95, respectively.
The enhancement factor defined by N$^{e^+e^-}_{measured}$/N$^{e^+e^-}_{hadronic\ sources}$ within
the mass range \linebreak 0.25 $<$ m $<$ 0.7 GeV/c$^2$ is 
2.6 $\pm$ 0.5 (stat.) $\pm$ 0.6 (syst.) 
for the '96 data.\hfill\break
\hskip -0.15cm
The pair yield as a function of the accompanying number of charged particles may be used as
a tool to disentangle different production mechanisms for the pairs. This 
dependence is shown in Fig.~\ref {result2} where the ratio between the measured and generator pair yield is shown
as a function of multiplicity.
Since the expectations from decays of produced hadrons grow
linearly with the multiplicity, a constant enhancement
indicates a linear dN$_{ch}$/d$\eta$ dependence.
Such 'normal' behaviour is displayed by the Dalitz decays dominating the two
lowest mass bins in Fig.~\ref {result2}, where we do not observe an enhancement.
In contrast, in the mass bin 0.25 MeV/c$^2$ $<$ m $<$ 0.68 MeV/c$^2$ where the strongest 
\begin{figure}[ht]
  \begin{flushleft}
\vspace{-1.5cm}    
    \begin{minipage}{0.5\linewidth}
      \epsfig{figure=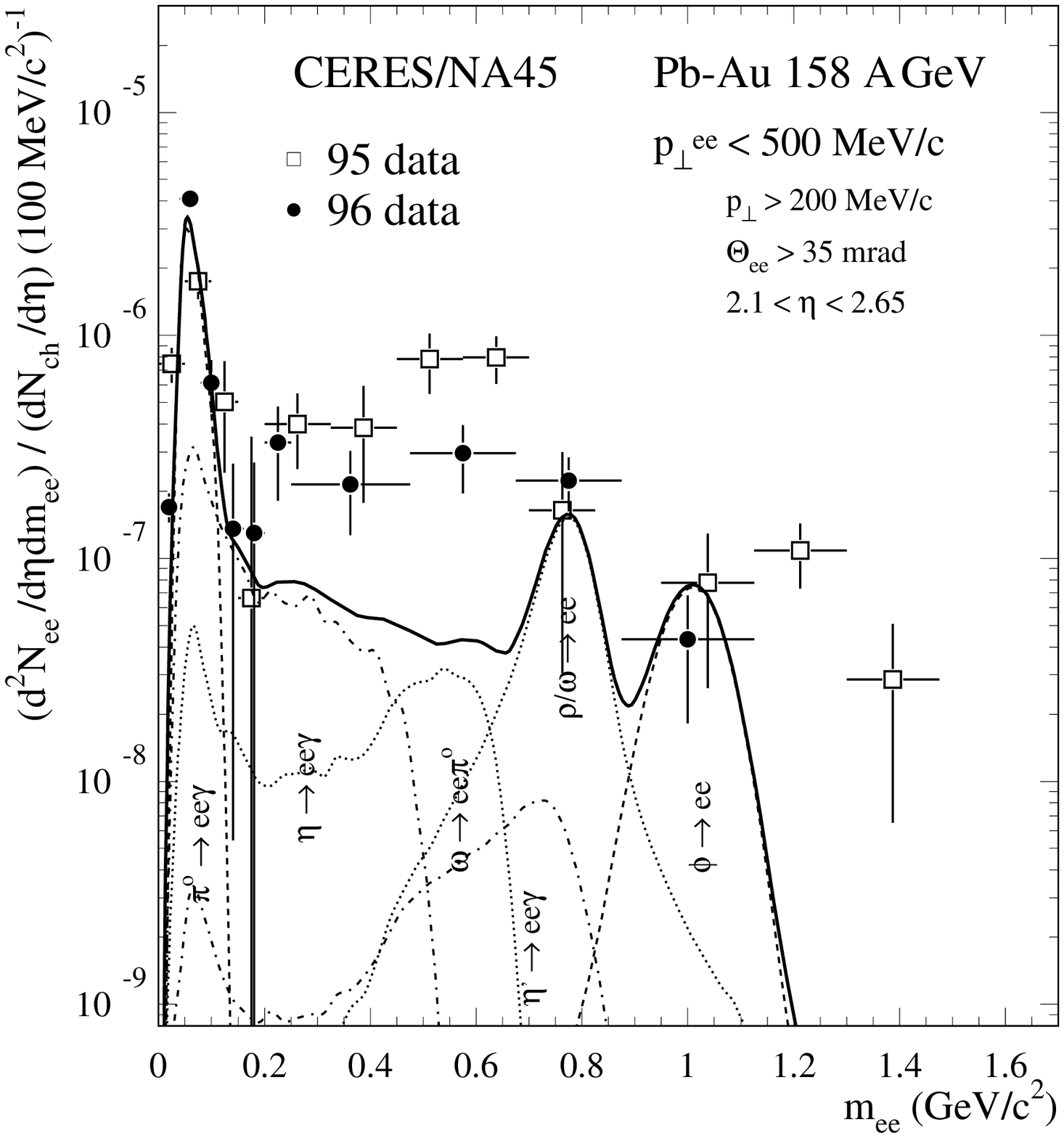,width=1.\linewidth}
    \end{minipage}
  \end{flushleft}
  \vspace{-10.3cm}
  \begin{flushright}
    \begin{minipage}{0.5\linewidth}
      \epsfig{figure=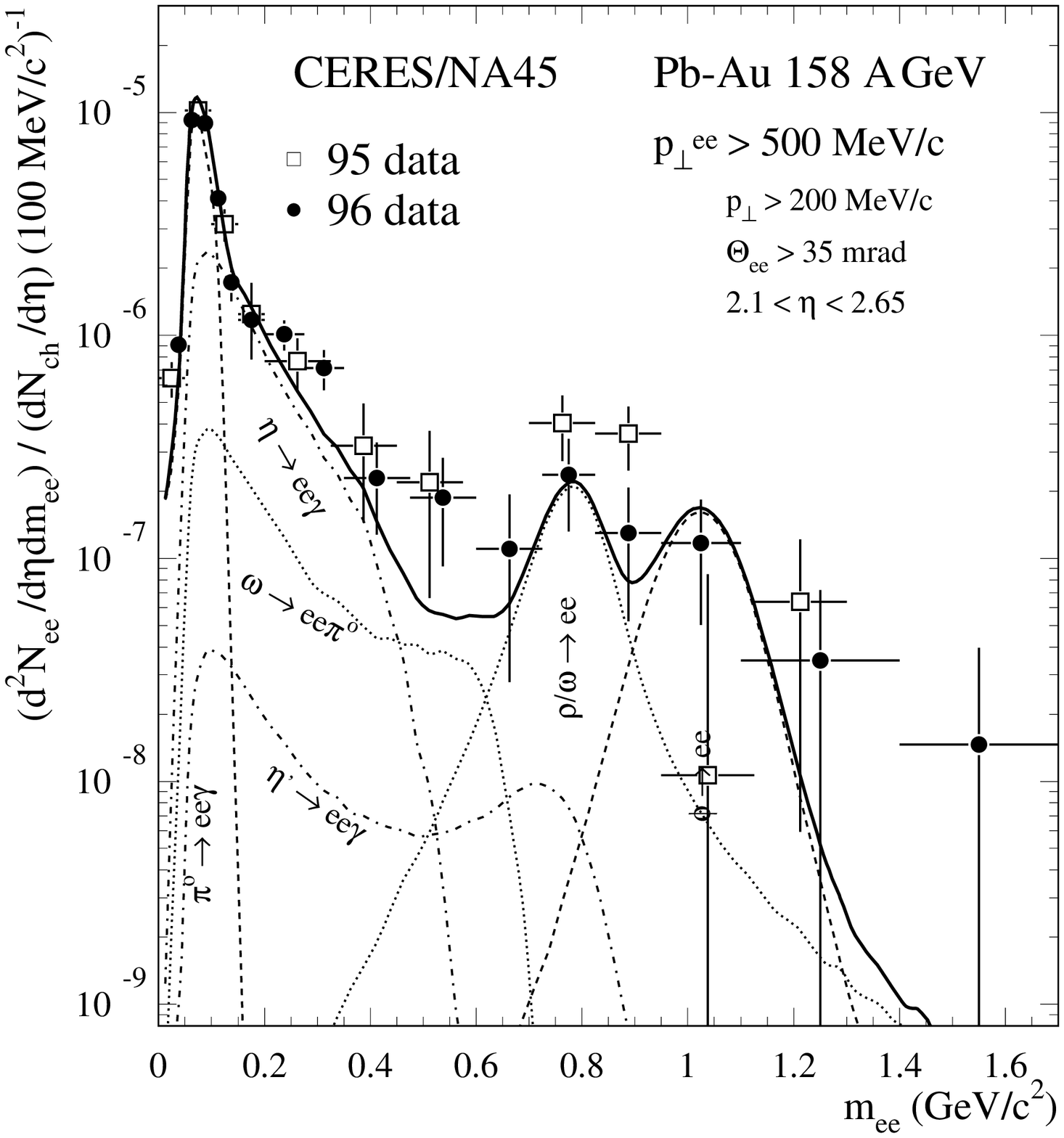,width=1.\linewidth}
    \end{minipage}
  \end{flushright}
\vskip -1cm  
\begin{minipage}{1.\linewidth}
  \begin{flushleft}
  \caption{\label{fig:ptrans}Inclusive e$^+$e$^-$ mass spectra 
  obtained from the $'$95 
  and $'$96 data analyses, se\-parated into samples with low (left) and high pair-p$_\perp$ (right).
  The data are compared to the expected contributions
  from hadron decays. The enhancement is most pronounced for low pair-p$_\perp$.}
  \end{flushleft}    
\end{minipage}
\end{figure}
\begin{figure}[ht]
  \begin{flushleft}
    \vskip -2.8cm  
    \begin{minipage}{\linewidth}
    \mbox{\hskip -4.cm\epsfig{file=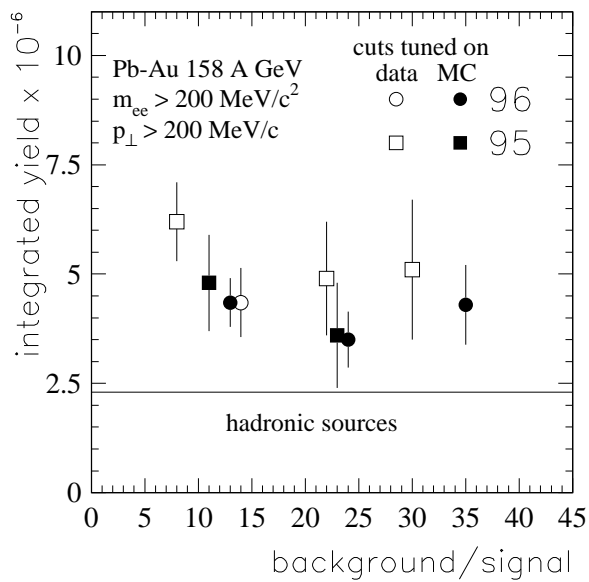,width=1\textwidth}}  
    \end{minipage}
  \end{flushleft}
  \begin{flushright} 
  \vskip -9.2cm  
    \begin{minipage}{7.cm}

    \caption{\label{starheaven}
    Integrated e$^+$e$^-$-pair yield as a function of the background-to-signal 
     ratio. Shown are the results of different analyses using different
     cut-tuning strategies, based either partially on data (open symbols) 
     or entirely on
     Monte Carlo simulation (full symbols).}

\end {minipage}
  \vskip 3.cm 
  \end{flushright}    
\end{figure}
enhancement is observed, the pair yield grows stronger than linear. \hfill\break
We also observe that the enhancement is most pronounced for pairs with
a small transverse momentum (see Fig.~\ref {fig:ptrans}). While for pairs with
p$_\perp^{ee}$ $<$ 500 MeV/c we observe an enhancement 5 $\pm$ 1.5 for the '96 data,
we find only a very modest enhancement under the 
complementary conditions. This enhancement is 
consistent with the small enhancement in the low-mass region reported by the muon
experiments \cite {phi/pi,helios-3} which have no acceptance for low p$_\perp$. 
\hfill\break
In comparison with the previously published data the new high-statistics 
analysis show a statistically consistent but systematically somewhat smaller yield. Therefore
we studied the stability of the result as a function of the background-to-signal
ratio for different cut-tuning strategies of the two data samples. The results
are summarized in Fig.\ref{starheaven}. 
First of all, the results of the '95 sample are seen to vary considerably between the two
choices of cut-tuning strategies,
whether one tunes the cuts
partially on the data (which was done historically) or with the MC (as described above).
These variations are 
much smaller in the '96 sample which is simply a fact of the larger statistics.
Much more important is, that the signal is stable and does not increase with the
B/S ratio. This makes very unlikely the hypothesis, that the observed enhancement is due to
a incorrectly subtracted background.\hfill\break
Another stability test we did was operating our spectrometer with different
polarities of the magnetic field. Both polarities show the same result within
the statistical errors.

\section{Search for direct photons}
Out of 2.5 million events we reconstructed 34 000 photons which converted into electron
pairs in the converter foil between the two silicon drift detectors. 
Due to the clear signature
shown in Fig.~\ref {trackpattern}, the background is negligible here 
\cite {matthias}. 
  \vskip -4.2cm
\begin{figure}[ht]
  \begin{flushleft}
    \begin{minipage}{0.6\linewidth}
    \mbox{\hskip 0.cm\epsfig{file=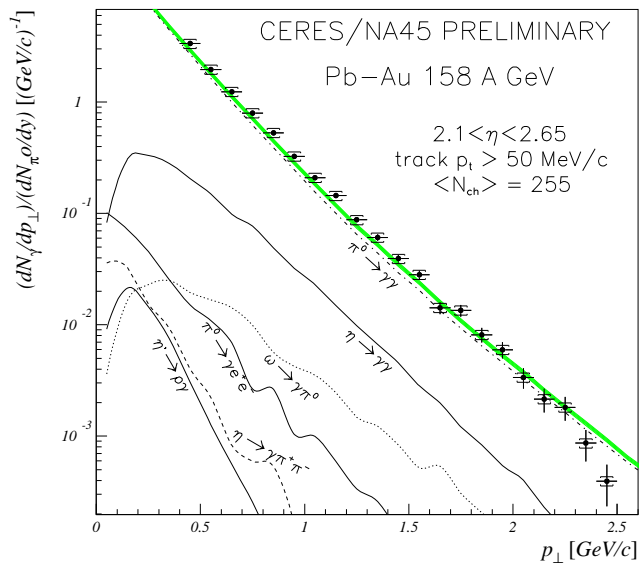,width=1\textwidth}}  
    \end{minipage}
  \end{flushleft}
  \begin{flushright} 
  \vskip -11.5cm  
    \begin{minipage}{6.cm}
    \caption{  \label {photon}
    Inclusive photon p$_\perp$ distribution in comparison with the expectations
    from hadron decays. The data and the generator are consistently normalized
    by the number of photons per $\pi^0$ with 
    p$_\perp$ $>$ 400 MeV/c within the 
    CERES acceptance. The shaded region indicates
    the systematic uncertainty of the
    generator, while the brackets indicate the systematic error of the data points.}
  \vskip 4.1cm  
\end {minipage}
  \end{flushright}  
\end{figure} 
\vskip -0.4cm
\hskip -0.4cm
Before obtaining the p$_\perp$ distribution
one has to correct the data for detector inefficiencies and momentum resolution. 
This correction is obtained with the same MC simulation
used for the low-mass pair analysis. The corrected p$_\perp$-distribution
is shown in Fig.~\ref {photon}. The lower bound of 400 MeV/c is due to the
poor efficiency for very low-momentum (large-deflection) tracks, while the
upper bound is due to limited statistics. In this interval the data closely follow
the expectations from hadron decays. It is the challenge of the photon
analysis to do the normalization of both yields as accurately as possible and
to minimize the systematic errors such that one is sensitive to an additional
source on the per cent level (corresponding to a low-mass pair enhancement
on top of the $\pi^0$-Dalitz peak). The normalization is done with respect to
the number of photons counted per $\pi^0$ in the same acceptance with a  
p$_\perp$ $>$ 400 MeV/c. Therefore we need to know the $\pi^0$-p$_\perp$-distribution
for our centrality and acceptance.  
This can be at best estimated from our measured h$^-$-spectrum
(as described below) by correcting it for contributions 
from $\bar{p}$, K$^-$ and secondaries from the $\eta$ $\rightarrow$ 3$\pi^0$
decay. In addition we have also investigated charged pion and uncharged pion distributions 
from other SPS experiments; these serve to estimate the systematical errors connected 
with our choice.\hfill\break
As indicated in Fig.~\ref {photon} the statistical errors of the measurement 
are negligible. More critical are the systematic errors due to
the Monte Carlo correction (6.3~\%), the uncertainty 
in the absolute normalization
(6.6~\%) and the uncertainty of the conversion length \linebreak (2.6~\%). The 
expectations from hadron decays also contain a systematic error originating
from the uncertainty in the p$_\perp$ distribution and the $\eta$/$\pi^0$
ratio (4~\%). \hfill\break
Within these errors we obtain a non-significant 
excess for photons with 
p$_\perp$ $>$ 400 MeV/c of 
(N$_{\gamma measur.}$-N$_{\gamma hadr.}$)/N$_{\gamma hadr.}$ = 12 \% $\pm$ 0.8\%$_{stat}$ 
$\pm$ 10.9\%$_{syst.}$. Within the errors this excess is 
constant as a function of
multiplicity. Further studies will verify this 
result with an independent data
sample taken with a different magnetic field, to reduce thereby the 
systematic uncertainties.

\section{Hadron analysis}
As indicated in Fig.~\ref {trackpattern}, we use two independent methods of
reconstructing charged hadrons \cite {fede}. First we can reconstruct low-momentum hadrons by
matching tracks in the SDD's to hits in the MWPC and determine the momentum by
the deflection between them. This method suffers from the large background
\begin{figure}[ht]
  \begin{flushleft}
 \vskip -2.6cm  
    \begin{minipage}{0.6\linewidth}
    \mbox{\hskip 0.cm\epsfig{file=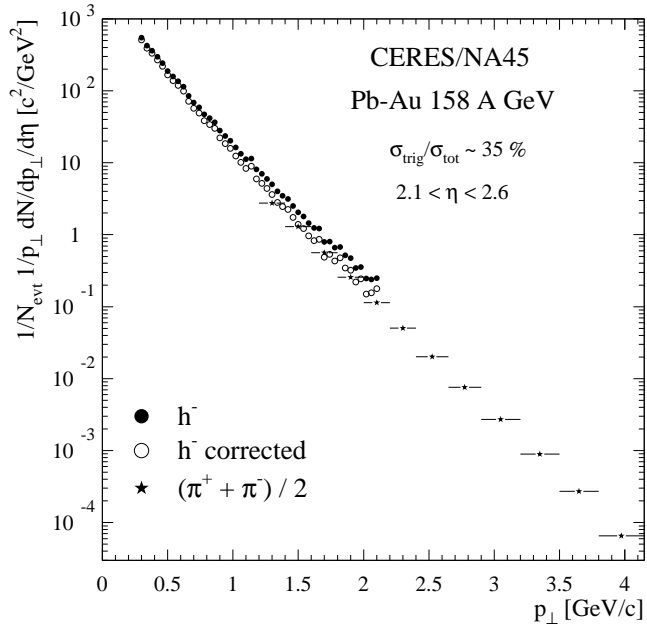 ,
     width=1\textwidth}}  
    \end{minipage}
  \end{flushleft}
  \begin{flushright} 
  \vskip -10.2cm  
    \begin{minipage}{6.cm}
    \caption{  \label {hadron}
    p$_\perp$-distributions for charged hadrons (labelled h$^-$) and identified
    charged pions (labelled ($\pi^+$+$\pi^-$)/2). The charged hadron spectrum
    is shown before (full circle) and after subtracting the contributions 
    from K$^-$ and anti-protons (open circle), the latter being in excellent
    agreement with the overlapping part of the pion spectra.}
\end {minipage}
  \vskip 1.4cm 
  \end{flushright}  
\end{figure}
(S/B = 2:1), which is determined by the event-mixing method.
The second method is sensitive to pions with p $>$ 4.5 GeV/c, since these
pions emit Cherenkov light. The Cherenkov ring radius is
used for the momentum determination.
The p$_\perp$-slopes of both methods show a very good agreement in the 
overlapping region (see Fig.~\ref {hadron}). In the region 
1.5 $<$ p$_\perp$ $<$ 3.5 GeV/c of the inclusive spectra, the inverse slope is
245 $\pm$ 5 MeV and increases by 2.4 \% with increasing centrality,
in the range of the 
most central 35\% of the geometric cross section. Identified $\pi^+$ and $\pi^-$ show
the same slope and the ratio $\pi^-$/$\pi^+$ = 1.030 $\pm$ 0.005 
for \linebreak 1 $<$ p$_\perp$ $<$ 2.4 GeV/c.
Compared to our previous pion analysis done for S-Au collisions
the inverse slope as a function of m$_\perp$ does not follow the systematics of the
Cronin effect, since the slope at small m$_\perp$ is much higher.
This is a hint for radial flow in Pb collisions already observed by 
other experiments \cite {rad_flow}.\hfill\break
We also studied anisotropic transverse flow by using only the SDD's in terms
of a Fourier decomposition of the azimuthal charged multiplicity distribution.
Our preliminary results show an anisotropic flow up to the 4th harmonic,
with the higher harmonics
having nearly the same strength than the first two.

\section{Conclusion and Outlook}
We have corroborated the low-mass pair enhancement previously observed by CERES.
This enhancement is most pronounced in the mass window from 
200 $<$ m $<$ 700 MeV/c$^2$,
at low p$_\perp$ and in events with a high charged particle multiplicity. In order to draw
final conclusions on whether these data exhibit a signal for the 
restoration of chiral symmetry
\cite {Rapp} better data are necessary. Therefore CERES has upgraded with a TPC installed
at the downstream end of the present spectrometer \cite {addendum}
to improve the signal-to-background ratio and the mass resolution. This
TPC was first tested in 1998 \cite {ana}. CERES will continue 
data taking in the 
next two years with higher statistics and at two different beam energies
(40 and 158 GeV per nucleon). \hfill\break
\\
The authors wish to thank the CERN staff for the good support.
We are grateful for the financial support by the German BMBF, the US DoE,
the MINERVA foundation, the Israel Science Foundation and the EC under
contract FMBICT972104.

\end{document}